\documentclass[epjST]{svjour}
\usepackage{graphics}
\usepackage{hyperref}
\usepackage{cite}
\usepackage{amsmath}
\usepackage{amssymb}
\usepackage{amsfonts}
\usepackage{appendix}
\usepackage{xcolor}
\RequirePackage{graphicx}

\begin{document}
\title{Thermodynamics of the quantum Schwarzschild black hole}
\author{
Leonardo Balart \inst{1}
\fnmsep 
\thanks{\email{leonardo.balart@ufrontera.cl}} 
\and 
Grigoris Panotopoulos \inst{1} 
\fnmsep 
\thanks{\email{grigorios.panotopoulos@ufrontera.cl}} 
\and 
\'Angel Rinc{\'o}n \inst{2} 
\fnmsep 
\thanks{\email{angel.rincon@ua.es}}
}
\institute{
Departamento de Ciencias F{\'i}sicas, Universidad de la Frontera, 
Casilla 54-D, 4811186 Temuco, Chile.
\and 
Departamento de Física Aplicada, Universidad de Alicante,
Campus de San Vicente del Raspeig, E-03690 Alicante, Spain.
}
\abstract{
We discuss some thermodynamic properties as well as the stability of a quantum Schwarzschild black hole, comparing the results with those obtained within a bumblebee gravity model. In particular, the Hawking temperature, $T_H$, the entropy, $S$,  the heat capacity, $C$, and the Gibbs free energy, $G$, are computed for both cases. In addition to that, we compute the Brown-York quasilocal energy and compare the solution with the  Schwarzschild case. We find that in both cases (quantum  Schwarzschild and bumblebee gravity model) the temperature, the entropy, and the heat capacity show the same functional form, under the replacement $\lambda^2 \rightarrow \ell$ and vice versa. 
Specifically, the temperature is found to be lower compared to the classical (Schwarzschild) solution, whereas the entropy is computed to be larger. Moreover, the heat capacity becomes more negative. Notably, a distinct contrast emerges in obtaining the Gibbs free energy between these two cases, and this distinction appears to stem from the ADM mass.
} 
\maketitle
%

\section{Introduction}
\label{intro}

In four-dimensional space-time, General Relativity (GR) stands as the sole viable theory of gravity capable of satisfying two fundamental requirements: i) diffeomorphism invariance, and ii) the strong equivalence principle \cite{DiCasola:2013iia}. Despite General Relativity's solid theoretical and experimental foundation, there exist compelling reasons to explore alternative theories of gravity.
Firstly, from a theoretical perspective, two of the most significant challenges that still exist in General Relativity are: 
i) the occurrence of singularities \cite{Penrose:1964wq,Hawking:1973uf}, and 
ii) the impossibility to achieve renormalization through standard processes of quantization \cite{Percacci:2017fkn}.
Secondly, from an observational point-of-view, the existence of dark sectors in the Universe highlights the crucial need for fundamental physics. To address the disparity between the ultraviolet (UV) and infrared (IR) sectors, incorporating new physics becomes essential in order to establish a connection.
This "new physics" could preserve the laws of gravity and might only necessitate the inclusion of new fields, which interact with ordinary matter through gravitational force.
Black holes (BHs) are of paramount importance both in classical and quantum regimes, which makes them particularly intriguing objects in the context of alternative theories of gravity. This notion is supported by a wealth of evidence, often summarized in the well-known "no-hair theorem" \cite{heusler_1996}.
Black holes exhibit several properties where classical and quantum effects coexist complexly. Among those phenomena, Hawking radiation \cite{hawking1,hawking2} holds a special place, as it has attracted a lot of interest over decades now, despite the fact that it has not been detected yet. It plays an important role in the multitude of effects observed in a black hole. 
Bekenstein-Hawking entropy is another manifestation of the quantum properties present in black holes. More specifically, Hawking's entropy unifies concepts from gravitation, thermodynamics, and quantum theory, and is therefore considered a gateway to the largely unexplored field of quantum gravity.
To be more precise, in the 1970s, Bekenstein \cite{Bekenstein:1972tm,Bekenstein:1973ur} and Hawking \cite{hawking1,hawking2} proved that black holes radiate as black bodies, with characteristic temperatures and entropies according to:
\begin{align}
    k T_{H} = \frac{\hbar \kappa}{2 \pi},   
    \hspace{2cm}  
    S_{H} = \frac{A_{hor}}{4 \hbar G_N}.
\end{align}
Here, $k$ is Boltzmann’s constant,
$\hbar = h/2\pi$  is Planck’s constant divided by 
$2\pi$, 
$\kappa$ represents the surface gravity, and $A_{hor}$ denotes the horizon's area. Those quantities seem to possess an intrinsically quantum-gravitational nature, as they rely on both Planck's constant and Newton's constant $G_N$. The concept of black hole thermodynamics has a remarkable impact on our understanding of the fundamental nature of black holes, connecting gravitational physics with quantum mechanics and thermodynamics. It also provides a potential link between the macroscopic world of general relativity and the microscopic world of quantum physics. However, a complete unification of these two theories, known as quantum gravity, is still an open problem.
Thermodynamics goes beyond temperature and entropy. Since black holes have thermal properties, it should be possible to apply the full set of thermodynamic principles, including concepts such as Carnot cycles and phase transitions, to systems involving black holes. Additional quite interesting black hole properties are, for instance: i) evaporation, ii) heat capacity, iii) phase transitions, and iv) thermodynamic volume, to name a few \cite{Carlip:2014pma,Wald:1999vt,Page:2004xp,Carlip:2008wv}.
Later on, a groundbreaking approach emerged to further develop the fundamental concepts of black hole thermodynamics. This new perspective, referred to as "black hole chemistry", redefines black hole mass as similar to chemical enthalpy, departing from the conventional notion of energy. In addition, the cosmological constant is reinterpreted as a thermodynamic pressure \cite{Kubiznak:2014zwa, Kubiznak:2016qmn, Frassino:2015oca}.

Black holes in general relativity have been studied extensively, mainly in the (1+3)-dimensional case, but also in (1+2)-dimensional version or extensions such as (1+4) (higher dimensional) gravity.

As already mentioned before, General Relativity has some problems, for example: i) the singularity problem and ii) the incompatibility with quantum mechanics. In this sense, it becomes natural that a modification of general relativity is needed.
In fact, we have striking examples of theories beyond General Relativity, for example, one of the most natural modifications is the well-known Brans-Dicke theory. In this theory,
Newton's constant is coupled to a scalar field which is non-minimally coupled to the Ricci scalar \cite{Brans:1961sx,Brans:1962zz,Cai:1996pj,Kang:1996rj,Liebling:1996dx}. 
Another relevant example is asymptotically safe (AS) gravity. Such a theory aims at obtaining a consistent and predictive quantum theory of the gravitational field \cite{Hawking:1979ig,Falls:2010he,Laporte:2021kyp}.
To summarize, there are several journeys available to promote classical gravitation to quantum gravity incorporating quantum effects if we become more flexible regarding the requirements of the theory. Some examples that could be mentioned here are the following
i) the improved formalism \cite{Bonanno:2000ep,Burschil:2009va,Koch:2013owa,Gonzalez:2015upa,Jusufi:2017vew,Moti:2018uvl,Pawlowski:2018swz,Moti:2018rho,Lu:2019ush,Rayimbaev:2020jye,Ishibashi:2021kmf,Lin:2022llz,Konoplya:2022hll,Mandal:2022stf,Ladino:2022aja,Atamurotov:2022iwj,Konoplya:2023aph,Lambiase:2023hng,Rincon:2020iwy}
ii) the variational parameter settings \cite{Koch:2014joa,Koch:2020baj,Laporte:2022wbu}
iii) scale-dependent gravity \cite{Koch:2016uso,Rincon:2017goj,Rincon:2018sgd,Rincon:2018lyd,Ovgun:2023ego,Rincon:2022hpy,Alvarez:2022wef,Alvarez:2022mlf,Panotopoulos:2021heb,Bargueno:2021nuc,Panotopoulos:2021tkk,Panotopoulos:2021obe,Rincon:2021hjj,Panotopoulos:2020mii,Rincon:2020cpz,Alvarez:2020xmk,Panotopoulos:2020zqa,Contreras:2019cmf,Fathi:2019jid}.
In the present article, we shall explore a newly discovered black hole that is inspired by loop quantum gravity \cite{Alonso-Bardaji:2021yls}. As claimed by the authors, that solution represents an asymptotically flat region, and it describes an exterior domain.
Additional studies related to quantum black holes within the framework of loop quantum gravity can be found in references, such as \cite{Modesto:2005zm, Modesto:2004xx, Zhang:2023yps, Perez:2017cmj, Liu:2020ola}.

In this work, our presentation is structured as follows: After this introductory section, we provide some contextual background on the quantum black hole model in Section \eqref{QBH_title}, where we introduce the metric and other necessary components to ensure the self-completeness of the article. Subsequently, in Section \eqref{termo}, we explore the black hole thermodynamics of this quantum model by calculating:
i) the Hawking temperature, denoted as $T_H$,
ii) the entropy, represented as $S$,
iii) the heat capacity, labeled as $C$, and
iv) the Gibbs free energy, identified as $G$.
In Section \eqref{bumblebee}, we conduct a comparison with Schwarzschild-like black holes in the bumblebee model. Then, in Section \eqref{QLE}, we compute the Quasilocal energy. The final section is dedicated to summarizing our findings.
Throughout this work, we shall be using the mostly positive metric signature $\{ -,+,+,+ \}$, and we employ geometrical units where $G_N=c=\hbar=1$.

\section{Quantum Black Hole and Metric Tensor} \label{QBH_title}

Loop quantum gravity predicts a quantized spacetime, potentially resolving the singularities that are present in General Relativity 
\cite{Ashtekar:2006uz}. In the same spirit, it is natural to expect a resolution of such singularities, or at least alleviate the situation \cite{Modesto:2006mx,Ashtekar:2005qt,Boehmer:2007ket}, within the framework of loop quantum gravity.
Nonetheless, a comprehensive quantum depiction of regions near a singularity remains elusive, necessitating the utilization of effective descriptions to incorporate quantum corrections. 
In order to keep the paper as self-complete as possible, we shall include some details about the novel space-time to be used along with this manuscript.

We start by rewriting a spherically symmetric space-time employing the Ashtekar new variables, see \cite{Boehmer:2007ket} for more details.   
Here, $E_i^a$ represents the set of triads, which are the canonical variables in LQG, and their corresponding SO(3) connections $A^i_a$.
In the spherically symmetric case only three pairs of canonical variables remain, namely $\{ \eta, P^{\eta}, A_{\phi}, E^{\phi}, A_x, E^x \}$. Note that a polar set of variables is chosen, and that $x$ is used since it is not necessarily described by the conventional radial coordinate \cite{Gambini:2008dy}. 
Alternatively, it has been shown that one may define the gauge invariant variables $K_i$ as a function of the connections $A_i$ and $\eta$, i.e. a more convenient variable can be defined. The latter implies that $E^x, K_x$ and $E^\phi, K_\phi$ represent the canonically conjugated pairs.
The diffeomorphism invariance of General Relativity is based on four constraints, the first of which is the Hamiltonian $\mathcal{H}$, while the other three ones result from the diffeomorphism constraint $\mathcal{D}$. For this particular case we can write down the quantities as follows:
\begin{align}
\begin{split}
    \mathcal{H}  = & -  \frac{\tilde{E}^\phi}{2\sqrt{\tilde{E}^x}} \left( 1+\tilde{K}_{\phi}^2 \right)
    - 2\sqrt{\tilde{E}^x} \tilde{K}_{x}\tilde{K}_{\phi} \ +
     \frac{1}{8\sqrt{\tilde{E}^x} \tilde{E}^\phi} \left( \left( \tilde{E}^x \right)^\prime \right)^2 
    \\
    &     
    - 
    \frac{\sqrt{\tilde{E}^x}}{2\left( \tilde{E}^\phi \right)^2} \left( \tilde{E}^x \right)^\prime 
    \left( \tilde{E}^\phi \right)^\prime +
    \frac{\sqrt{\tilde{E}^x}}{2 \tilde{E}^\phi } \left( \tilde{E}^x \right)^{\prime \prime},
\end{split}
    \\
    \mathcal{D}   =  &\ \ \  \left( \tilde{E}^x \right)^\prime \tilde{K}_{x}+ \tilde{E}^\phi \left( \tilde{K}_{\phi}\right)^\prime.
\end{align}
The meaning of the notation is as follows: 
i) the prime represents the derivative with respect to $x$, 
ii) $\tilde{E}^i$ are the symmetry reduced triad components, and finally, 
iii) $\tilde{K}_i$ represent the conjugated momenta with $i=\left\lbrace x, \phi \right\rbrace$.
Note that the holonomies of the connection have well-defined operators in loop quantum gravity. This is the reason why a polymerization procedure is needed (see \cite{Gambini:2021uzf} for more details).
Roughly speaking, the idea is to replace the variables by an exponential form. 
In the case where real variables are involved, a convenient replacement is of the form 
\begin{align}
\tilde{x} &\rightarrow \frac{\sin\left( \lambda x \right)}{\lambda}
\end{align}
where the last parameters have the usual meaning, i.e. i) $x$ is the variable and ii) $\lambda$ is the polymerization parameter. For the construction, the classical theory is recovered taking the limit
$\lambda \to 0$. It should be noted that $\lambda$, i.e. the polymerization parameter, is directly related to the length of the loop along which the holonomy is computed, since this is responsible for the space-time discretization.
It is important to emphasize that anomalies might occur when using the previous substitution, as the modified constraint algebra typically does not close.
Alternatively, both $K_\phi$ and $E^\phi$ can be replaced by $\tilde{K}_\phi$ and $\tilde{E}^\phi$ via the following changes:
\begin{align}
    \tilde{K}_\phi & \rightarrow \frac{\sin\left( \lambda K_\phi \right)}{\lambda}
    \\
    \tilde{E}^\phi & \rightarrow \frac{E^\phi}{\cos \left( \lambda K_\phi \right)}
\end{align}
and thus a theory free of anomalies \cite{Alonso-Bardaji:2021yls} may be obtained. 
On the one hand, the canonical transformation is bijective 
while $\cos \left( \lambda K_\phi \right) \neq 0$ and the dynamical content of the theory is equivalent to that obtained in GR.
On the other hand, the case $\cos \left( \lambda K_\phi \right) =0$ may be physically relevant (i.e., introduce new physics). Since the Hamiltonian constraint diverges there, a regularization is considered.

The procedure for obtaining the concrete space-time can be consulted, as mentioned before, in \cite{Alonso-Bardaji:2021yls}, but at this point it is clear that just taking a simplified chart 
$\{t,x\} = \{\tilde{t}, \tilde{r}\}$ and setting $E^x = \tilde{r}^2$ and $K_{\phi} = 0$ we are able to obtain the appropriate space-time.
In what follows, we shall introduce an effective quantum-corrected Schwarzschild space-time originally discussed in Ref.~\cite{Alonso-Bardaji:2021yls,Alonso-Bardaji:2022ear}. As the discussion from the mathematical point of view is too involved, we shall spear the technical details here, and we refer the interested reader to ~\cite{Alonso-Bardaji:2021yls,Alonso-Bardaji:2022ear}.

Assuming the most general form of a static, spherically symmetric line element, we can write down the following expression
 
\begin{equation}
ds^2 = - f(r) dt^2 + h(r)^{-1} dr^2 + r^2 d\Omega^2
\,\,\label{metric-f} \,  .
\end{equation} 
characterized by two different metric functions, where 
\begin{equation}
d\Omega^2 = d \theta^2 + \sin^2 \theta \: d \phi^2
\,\,\label{} \,  
\end{equation}
is the metric tensor of the two-dimensional unit sphere, while the metric functions are found to be
\begin{equation}
f(r) = 1 - \frac{2 m}{r}
\,\,\label{} \,  
\end{equation} 
\begin{equation}
h(r) = g(r) f(r)
\,\,\label{} \,  
\end{equation} 
\begin{equation}
g(r) = 1 - \frac{r_0}{r}
\,\,\label{} \, . 
\end{equation} 
where the parameter $r_0$, with dimensions of length, is due to quantum effects, and it is given by
\begin{equation}
r_0 = 2 m \Bigg( \frac{ \lambda^2 }{1 + \lambda^2} \Bigg)
\,\,\label{} \,  .
\end{equation} 
Regarding the radial metric function, $h(r)$, the algebraic equation $h(r)=0$ admits two roots located at $r_+ = 2 m$ and $r_- = r_0$.
Please take note that the parameter $m$ cannot be identified with the mass of the black hole. This distinction arises due to the existence of various geometric definitions of mass in this scenario, none of which are, in general, mutually interchangeable.

\section{Thermodynamic Quantities} \label{termo}

Here we consider two ways of finding the Hawking temperature of the black hole under consideration. First, we can calculate the temperature using the expression for surface gravity, that is 
\begin{equation}
T_H = \frac{\kappa}{2 \pi}  = \frac{1}{4 \pi}  \frac{\partial_r f(r_+)}{\sqrt{f(r_+)h(r_+)^{-1}}}
= \frac{1}{8 \pi m  }\frac{1}{\sqrt{1+\lambda ^2}}
\,\,\label{th-1} \,  .
\end{equation} 
We can also compute the Hawking temperature if we consider the expression based on the tunneling formalism given in Ref.~\cite{Angheben:2005rm}
\begin{equation}
T_H = \frac{1}{4 \pi} \sqrt{\partial_r f(r_+)\partial_r h(r_+)}
= \frac{1}{8 \pi m  }\frac{1}{\sqrt{1+\lambda ^2}}
\,\,\label{th-2} \,  .
\end{equation} 
Notice that we can write
\begin{equation}
T_H = \frac{1}{8 \pi m } \left(1-\frac{\lambda ^2}{2}+\frac{3 \lambda ^4}{8}+\mathcal{O}(\lambda^6)\right)
\,\,\label{} \,  
\end{equation} 
and we can see that in the classical limit the Hawking temperature of the Schwarzschild black hole $T_0 \equiv 1/(8 \pi m)$ is recovered.

\begin{figure*}[ht!]
\centering
\includegraphics[scale=0.680]{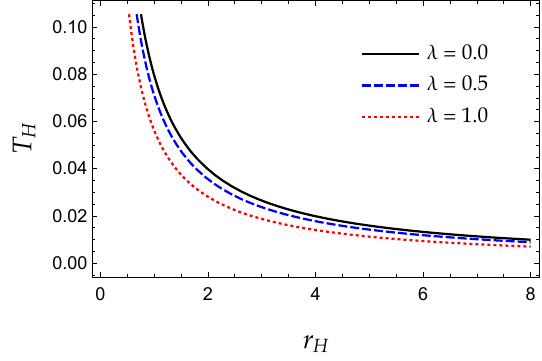} \
\includegraphics[scale=0.680]{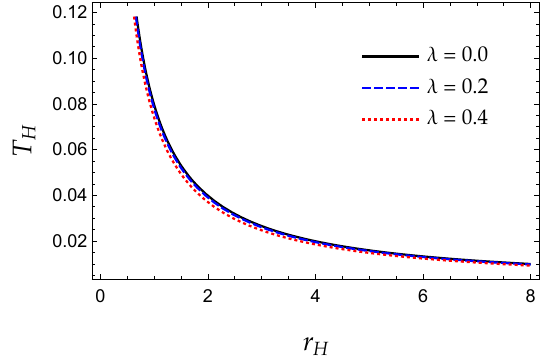} \
\\
\includegraphics[scale=0.680]{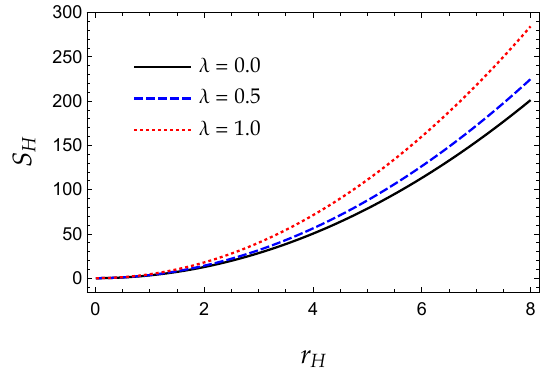} \
\includegraphics[scale=0.680]{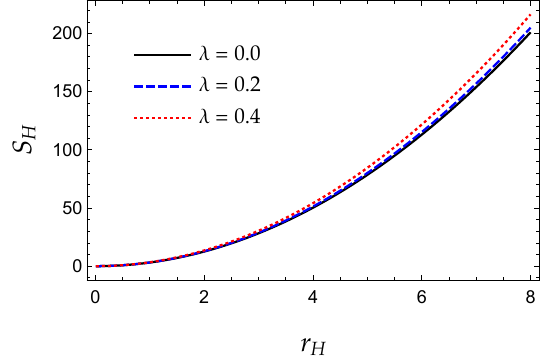} \
\caption{
Hawking temperature, $T \equiv T_H$, and Bekenstein-Hawking entropy, $S \equiv S_H$, for the quantum black hole for different values of the parameter $\lambda$.
{\bf{First row:}} Exact Hawking temperature against the horizon (left) and Approximated Hawking temperature against the horizon (right).
{\bf{Second row:}} Exact Bekenstein-Hawking entropy against the horizon (left) and Approximated Bekenstein-Hawking entropy against the horizon (right).
}
\label{fig:1} 	
\end{figure*}

\begin{figure*}[ht!]
\centering
\includegraphics[scale=0.680]{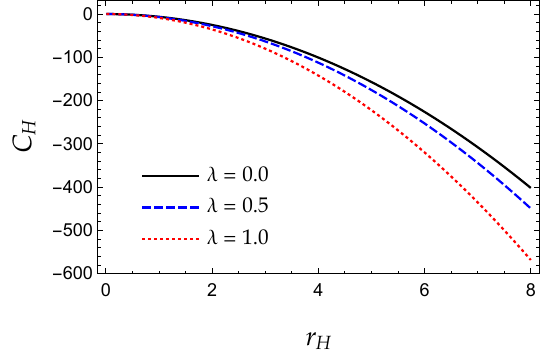} \
\includegraphics[scale=0.680]{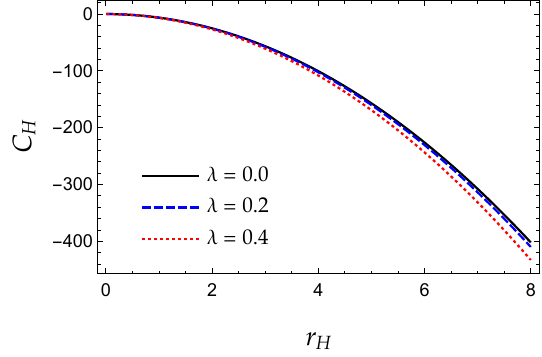} \
\\
\includegraphics[scale=0.680]{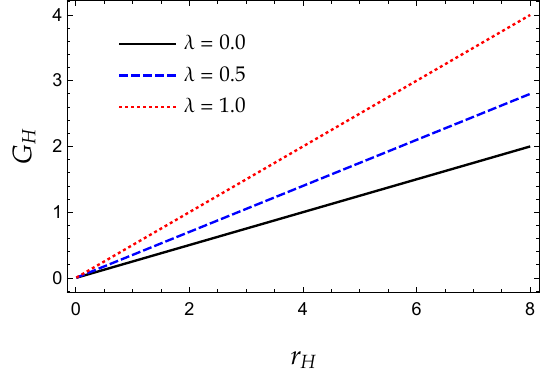} \
\includegraphics[scale=0.680]{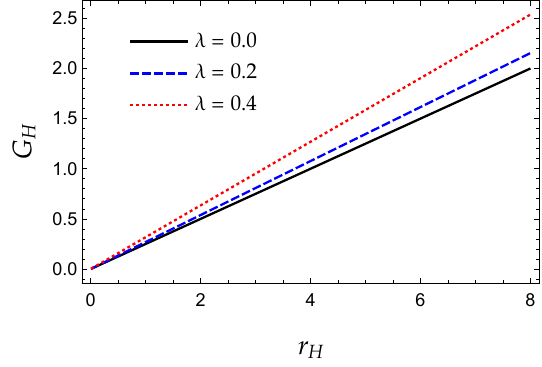} \
\caption{
Heat Capacity, $C \equiv C_H$, and Gibbs free energy, $G \equiv G_H$, for the quantum black hole for different values of the parameter $\lambda$. 
{\bf{First row:}} Exact Heat capacity against the horizon (left) and Approximated Heat capacity entropy against the horizon (right).
{\bf{Second row:}} Exact Gibbs free energy against the horizon (left) and Approximated Gibbs free energy against the horizon (right).
}
\label{fig:2} 	
\end{figure*}
 
From the temperature, we can obtain the entropy
\begin{equation}
S = \int \frac{dm}{T_H} = \pi  r^2_+ \sqrt{1+\lambda ^2}
\,\,\label{Ent} \,  ,
\end{equation} 
where $S_0 \equiv \pi r^2_+$ is the entropy of the Schwarzschild black hole.

As is the case with the Hawking temperature, we can perform an expansion around small values of $\lambda$ to examine how the classical solution is altered. In other words, we can analyze the modifications to the classical solution by expanding with respect to small $\lambda$ values. Thus we have
\begin{align}
    S = \pi r_{+}^2 
    \bigg( 
   1 + \frac{\lambda ^2}{2}  -\frac{\lambda ^4}{8} + \mathcal{O} \left(\lambda ^6\right)
    \bigg) .
\end{align}
The heat capacity is calculated with the following expression
\begin{equation}
C =  T_H \left(\frac{\partial S}{\partial T_H}\right) = T_H \frac{\left(\frac{\partial S}{\partial r_+}\right)}{\left(\frac{\partial T_H}{\partial r_+}\right)}
\,\,\label{H-C} \,  ,
\end{equation} 
from which it follows
\begin{equation}
C =  -2 \pi  r^2_+ \sqrt{1+\lambda ^2} 
\,\,\label{H-C} \,  .
\end{equation}
Note that $C_0 = -2 \pi r^2_+$ is the heat capacity of Schwarzschild solution.
As occurred in the previous expression, we can expand for small values of $\lambda$ and we obtain
\begin{align}
    C &=  -2 \pi r^2_{+} 
    \bigg( 
   1 + \frac{\lambda ^2}{2}  -\frac{\lambda ^4}{8} + \mathcal{O} \left(\lambda ^6\right)
    \bigg) .
\end{align}

The Gibbs free energy is 
\begin{equation}
G =  M_{ADM} - T_HS = \frac{r_+}{2} \left(\frac{3}{2}-\frac{1}{1+\lambda ^2}\right)
\,\,\label{G-E} \,  .
\end{equation}
In the limit $\lambda \rightarrow 0$, $G$ is recovered for the Schwarzschild black hole, which is $G_0 \equiv r_{+}/4$. As we proceed before, taking an expansion around a small value of $\lambda$ we find the expression:
\begin{align}
    G &= \frac{r_{+}}{4}
    \bigg(
     1 + 2 \lambda ^2 -2 \lambda ^4
    + \mathcal{O}(\lambda^6)
    \bigg)   \,  ,
\end{align}
where the ADM mass of the asymptotically flat spacetime is given by (see appendix for further details):
\begin{equation}
M_{ADM} = \frac{1 + 2\lambda^2}{1 + \lambda^2} m
\,\,\label{ADM-Q} \,  .
\end{equation} 
The correction coming from the weak-$\lambda$ contribution is given as
\begin{align}
    M_{ADM} &= m \Bigl( 1 + \lambda^2 - \lambda^4 + \mathcal{O}(\lambda^5)\Bigl) \,  .
\end{align}

In this case, we do not have a standard Smarr formula since the mass of Komar (see for example Ref.~\cite{Heusler:1994wa}) for the model presented in Eq.~(\ref{metric-f}) is
\begin{equation}
M_K = m \sqrt{1-\frac{r_0}{r}}
\,\,\label{} \,  .
\end{equation}

If we evaluate it in $r_+$ we get
\begin{equation}
M_H = \frac{m}{\sqrt{1+\lambda ^2}} 
\,\,\label{massfinal} \,  .
\end{equation}
In this way, $M_H \neq 2 T_HS $ and rather it is fulfilled that
\begin{equation}
M_H = \frac{2}{\sqrt{1+\lambda^2}}  T_H S
\,\,\label{Smarr-Q} \,  .
\end{equation}
To make the last expression more transparent, we invite the reader to check if: 
i) we use \eqref{th-2} and \eqref{Ent} to construct the product $T_H S$, and
ii) we use \eqref{massfinal} as the value of $M_H$, 
we can remove $m$ and trivially identify the effective Smarr law for this quantum Schwarzchild black hole.

We can notice that quantum effects lead to different outcomes when calculating the mass of the system using either the Komar formulation or the ADM mass, unlike what happens classically. In the case of the ADM mass, as seen in Eq.~(\ref{ADM-Q}), we note an increase in the total gravitational contribution compared to the classical scenario, a ``gravitational anti-screening'' effect. Conversely, when considering the Komar integral, we find that the Komar mass matches the classical mass $m$ as $r\rightarrow \infty$. However, at the event horizon, the result is lower than that of the classical case, as shown in Eq.~(\ref{massfinal}), a ``gravitational screening'' effect. Additionally, in the Smarr formula that we have derived, the result of the product of temperature and entropy yields the same value as in the classical case. However, as we mentioned, the Komar mass at the event horizon decreases compared to its classical counterpart.

\section{Comparison with Schwarzschild-like black holes in bumblebee model} \label{bumblebee}

A bumblebee gravity model is an extended gravitational model including Lorentz-violating terms. The model is described by the action \cite{Casana:2017jkc}
\begin{equation}
I_B = \int d^4x\mathcal{L}_B
\,\,\label{S-Bum} \,  ,
\end{equation}
where the Lagrangian density is given by
\begin{align}
\begin{split}
\mathcal{L}_B = \sqrt{-g} 
\Bigg(
\frac{R}{16 \pi} &+ \frac{1}{16 \pi} \xi B^\mu B^\nu R_{\mu\nu} 
- \frac{1}{4} B_{\mu\nu} B^{\mu\nu} - V(B^\mu) + \mathcal{L}_M
\Bigg)
\,\,\label{L-Bum} \,  ,
\end{split}
\end{align}
with $\mathcal{L}_M$ being the Lagrangian density of matter content, the vector field $B_\mu$ is the Lorentz-violating bumblebee field, while $B_{\mu\nu} = \partial_\mu B_{\nu} - \partial_\nu B_{\mu}$ is the corresponding field strength, and $\xi$ is the coupling of $B_\mu$ to gravity. Finally, the potential $V$ is chosen such that it provides a non-vanishing vacuum expectation value for $B_\mu$, which could have the following general functional form \cite{Casana:2017jkc}
\begin{equation}
V = V(B_\mu B^\mu \pm b^2) 
\end{equation}
where $b^2$ is a positive real constant. 

For vacuum solutions, where the matter stress-energy tensor vanishes, and assuming that the vector field remains frozen in its vacuum expectation value, $B_\mu = b_\mu$,
the solution to the field equations for the metric tensor is found to be \cite{Casana:2017jkc}
\begin{equation}
ds^2 = - f(r) dt^2 + h(r)^{-1} dr^2 + r^2 d\Omega^2
\,\,\label{} \,  .
\end{equation} 
where 
\begin{equation}
f(r) = 1 - \frac{2 m}{r} ,
\,\,\label{} \,  
\end{equation} 
\begin{equation}
h(r) = g(r) f(r) ,
\,\,\label{h-bumb} \,  
\end{equation} 
and
\begin{equation}
g(r) = (1 + \ell)^{-1}
\,\,\label{g-bumb} \, .  
\end{equation} 
where the notation $\ell = \xi b^2$ is used.

Following Ref.~\cite{Gomes:2018oyd} we can obtain the same thermodynamic quantities as above. The Hawking temperature in this case is computed to be
\begin{equation}
T_H = \frac{1}{8 \pi m  }\frac{1}{\sqrt{1+\ell}}
\,\,\label{t-b} \,  ,
\end{equation} 
and in the approximation $\ell \ll 1$ we get
\begin{equation}
T_H = \frac{1}{8 \pi m } \left(1-\frac{\ell}{2}+\frac{3 \ell^2}{8} + \mathcal{O}(\ell^3)\right).
\,\,\label{} \,  
\end{equation} 
From the temperature we can obtain the entropy
\begin{equation}
S = \int \frac{dm}{T_H} = \pi  r^2_+ \sqrt{1+\ell}
\,\,\label{} \,  .
\end{equation} 

The heat capacity is 
\begin{equation}
C =  -2 \pi  r^2_+ \sqrt{1+\ell} 
\,\,\label{H-C-B} \,  ,
\end{equation}

In this case the modifications are due to a matter-bumblebee coupling term unlike what happens with the quantum Schwarzschild black hole.

Unlike the quantum Schwarzschild model, in this case the metric is not asymptotically flat, so the definition for computing the ADM mass does not apply \cite{Xu:2022frb}.

From the Ref.~\cite{Heusler:1994wa} or \cite{Xu:2022frb} we can get the Komar mass
\begin{equation}
M_K = M_H = \frac{m}{\sqrt{1+\ell}}
\,\,\label{Komar-B} \,  .
\end{equation} 
So here we do not find a standard Smarr formula either, but rather the following expression
\begin{equation}
M_H = \frac{2}{\sqrt{1+\ell}}  T_H S
\,\,\label{Smarr-B} \,  .
\end{equation} 

Again, as we observed for Eq.~(\ref{Smarr-Q}), here also when calculating the product of entropy and mass we obtain the same value as in the classical case. However, the Komar mass at the horizon is smaller than its classical counterpart. 
In this case, unlike the quantum Schwarzschild black hole, we are dealing with an extended gravitational model based on the coupling of a bumblebee vector field to gravity. 
The parameter $l$ is related to  the constant coupling of the vector field and its magnitude is associated with the effects of Lorentz symmetry violation in nature. Consequently, the observed decrease in the Komar mass evaluated at the horizon, compared to the classical case, represents a  manifestation of Lorentz symmetry breaking.

Both expressions~(\ref{Smarr-Q}) and~(\ref{Smarr-B}), derived at the event horizon, depend similarly on the parameters $\lambda$ and $l$ in their respective models. However, it is important to highlight the following distinctions: while the quantum Schwarzschild black hole originates from the quantization of spacetime, the other model arises from the coupling of a bumblebee field to gravity. Moreover, the fact that metric for the bumblebee model is not asymptotically flat reflects that the Komar mass does not coincide with $m$ as $r\rightarrow \infty$, unlike the previous model. That is, if we assume the Komar mass as the total mass, then in the first case an observer very far from the quantum black hole will not perceive quantum effects if the Komar integral is considered to measure them, unlike what would be perceived if the bumblebee model is considered, as we can see from eq.~(\ref{Komar-B})

\section{Quasilocal Energy} \label{QLE}

In addition to the identities that can be obtained in the context of the laws of BH thermodynamics, there is another identity that also relates quantities evaluated at the event horizon and at infinity which is based on the Brown-York quasilocal energy~\cite{Bose:1998uu}, namely
\begin{equation}
E(r_h) - E(\infty) =  M_H 
\,\,\label{QLE-rel} \,  .
\end{equation}
Here $M_H$ is the Komar mass evaluated at $r_+$ and the expression for the Brown-York quasilocal energy is given by~\cite{Brown:1992br}
\begin{equation}
E(r) = \frac{1}{8 \pi}\int_{B} (k-k_0) \sqrt{\sigma} \,  d^2 x
\,\,\label{QLE-def} \,  ,
\end{equation}
where $B$ is the 2-dimensional spherical surface, $k$ is the trace of the extrinsic curvature of $B$, $\sigma$ is the determinant of the metric of $B$ and $k_0$ is a reference term.

If we consider the metric element given in Eq.~(\ref{metric-f}) and choose a Minkowski spacetime as a reference, we obtain that quasilocal energy for a $r\geq r_+$ is
\begin{equation}
E(r) = r - r \sqrt{h(r)}
\,\,\label{} \,  .
\end{equation}
In particular, for the quantum Schwarzschild black hole we get
\begin{equation}
E(r) = r-\sqrt{\frac{(r - 2 m) \left[r+(r -2 m) \lambda ^2\right]}{1+\lambda ^2}}
\,\,\label{}
 \,  ,
\end{equation}
and
\begin{equation}
E(r_+) = r_+ = 2 m
\,\,\label{E-hor}
 \,  ,
\end{equation}
\begin{equation}
E(\infty) = \frac{1 + 2 \lambda ^2}{1+\lambda ^2}m
\,\,\label{E-inf} \,  .
\end{equation}
Note that $E(\infty)$ is equal to $M_{ADM}$ as expected for an asymptotically flat spacetime according to Refs.~\cite{Brown:1992br,Brewin:2006qe}.

From Eqs.~(\ref{E-hor}) and~(\ref{E-inf}) it can be obtained that
\begin{equation}
E(r_+) - E(\infty) = \frac{m}{1+\lambda ^2} .
\end{equation}

As we previously obtained the Komar mass on the horizon $r_+$ is
\begin{equation}
M_H = \frac{m}{\sqrt{1+\lambda ^2}} 
\,\,\label{} \,  .
\end{equation}

From the above, the following relation is obtained
\begin{equation}
\sqrt{1+\lambda ^2}\, (E(r_+) - E(\infty)) =  M_H 
\,\,\label{QLE-rel-mod} \,  ,
\end{equation}
which differs from the relationship given in Eq.~(\ref{QLE-rel}). 
However, the quantum black hole model satisfies a more general identity given in Ref.~\cite{Balart:2009xr}. When $g_{00} \neq -(g_{11})^{-1}$, $g_{22} = r^2$ and $g_{33} = r^2 \sin^2{\theta}$ the following identity can be written
\begin{equation}
E(r_+) - E(\infty) = e^{-\delta(r_+)} M_H 
\,\,\label{QLE-rel-gral} \, .
\end{equation}
Here, if we adopt the notation of the line element provided by Eq.~(\ref{metric-f}) then 
\begin{equation}
e^{2 \delta(r_+)} = \frac{f(r_+)}{h(r_+)}
\,\,\label{} \, .
\end{equation}

For the quantum Schwarzschild model it is found that
\begin{equation}
e^{- \delta(r_+)} = \frac{1}{\sqrt{1+\lambda ^2}} 
\,\,\label{} \, .
\end{equation}
Thus in this case relation~(\ref{QLE-rel-gral}) is equivalent to relation~(\ref{QLE-rel-mod}).

In order to compute the quasilocal energy for the bumblebee model we consider as reference an asymptotically non-flat spacetime determined by the metric function~(\ref{h-bumb}) with~(\ref{g-bumb}). Thus we obtain
\begin{equation}
E(r) = r \left(\sqrt{\frac{1}{1+\ell}}- \sqrt{\frac{1 - 2m/r}{1+\ell}} \right)
\,\,\label{E-BY-B}
 \,  ,
\end{equation}
and from here follows 
\begin{equation}
E(r_+) =\frac{2 m}{ \sqrt{1+\ell}}
\,\,\label{}
 \,  ,
\end{equation}
\begin{equation}
E(\infty) =\frac{m}{ \sqrt{1+\ell}}
\,\,\label{E-inf-B}
 \,  .
\end{equation}
Therefore the identity~(\ref{QLE-rel}) is satisfied in this case.

From Eq.~(\ref{E-BY-B}) we can calculate the total energy of the black hole, that is, we can evaluate the quasilocal energy in $r \rightarrow\infty$ to obtain the physical mass. Usually when the total energy of the system is calculated under the Brown-York prescription, to remove divergences its value must be normalized with a reference spacetime. In cases where spacetimes that are asymptotically non-flat are considered, one chooses a non-flat geometry as a reference to obtain a finite value for the quasilocal energy. In particular for the black hole model coupled with a bumblebee field, we consider the functions $g_{tt}$ and $g_{rr}$ defined in Eqs.~(\ref{h-bumb}) and~(\ref{g-bumb}) as non-flat geometry by setting $m = 0$. It can be checked in this case that if the reference geometry is not considered or if a flat spacetime is set as reference, a divergence is obtained. Other examples of asymptotically non-flat spacetime are found in the literature, one of them is the black hole with a global monopole (e.g. see~\cite{Bose:1998uu}), where if the mass is calculated with the Brown-York prescription, one also finds a mass scaled by factor that depends on a constant parameter of the model. Without being the same conceptually or mathematically, in the case of the field bumblebee coupled to the black hole there is also a global effect on spacetime, as seen in Eq.~(\ref{E-inf-B}).

On the other hand, in the quantum black hole, where the reference geometry is flat, we obtain the mass~(\ref{E-inf}), so at first glance we could consider that the quantum effect that manifests itself with the presence of the parameter $\lambda$ is a global effect, as in the case of gravitation with bumblebee field. But in this case it should rather be interpreted as a quantum correction to the mass of the black hole. So here one could think that the quantum effect that manifests by the parameter $\lambda$ is intrinsic to the presence of a mass.

\section{Conclusions}

Summarizing our work, we have explored the black hole thermodynamics of a novel quantum black hole model inspired by loop quantum gravity in a four-dimensional space-time with a vanishing cosmological constant.
Specifically, we have investigated the alterations in temperature, entropy, heat capacity, and Gibbs free energy compared to their classical counterparts (when $\lambda = 0$). Furthermore, we recalculated these quantities in the context of bumblebee gravity for the purpose of comparison. Notably, we observed that the temperature, entropy, and heat capacity exhibit similarity under the replacement $\lambda^2 \rightarrow \ell$.
Additionally, we noted a discrepancy in obtaining the Gibbs free energies between the quantum black hole scenario and the bumblebee gravity case. A difference is also presented between the form of the identity in the context of quasilocal energy that each model satisfies.
These discrepancies are related to the asymptotic behavior of each spacetime.

Moreover, it should be mentioned that this quantum black hole model affects the thermodynamic quantities in a non-trivial way. For example, note that the Hawking temperature is corrected by the factor $1/\sqrt{1+\lambda^2}$, which implies that in this case $T_H$ is lower compared to the classical case. 
So such a change in the Hawking temperature could suggest that the mass should be rescaled by this factor (i.e, $m \rightarrow m \sqrt{1+\lambda^2}$), but after calculating $M_H$ (see \eqref{massfinal}) it becomes clear that quantum corrections, although subtle, modify the thermodynamic parameters in a different manner, since $M_H = m/\sqrt{1+\lambda^2}$ instead of $M_H = m \sqrt{1+\lambda^2}$.
Also note that the ADM mass is increased by the additional term $\lambda^2/(1+\lambda^2)m$, which is absent in the classical solution. More importantly, the prefactor of the ADM mass is $(1+2\lambda^2)/(1+\lambda^2)$, making this correction significantly non-trivial.

Finally, as far as the quasilocal energy is concerned, the basic identity $E(r_h)-E(\infty) = M_H$ is satisfied in the case of the bumblebee gravity model, where the geometry is asymptotically non-flat. On the contrary, in the case of the quantum black hole, which is asymptotically flat, the basic identity is not satisfied, a fact that may be interpreted as a quantum correction to the mass of the black hole.


\section*{Appendix}

From Ref.~\cite{Wald} we can use the following expression for the ADM mass
\begin{equation}
M_{ADM} = \frac{1}{16 \pi} \lim\limits_{r\rightarrow \infty} \int \sum_{i,j = 1}^{3} (g_{ij,i} - g_{ii,j}) n^j dA
\,\,\label{} \, ,
\end{equation}
where the integral is over a sphere of constant $r = \sqrt{(x^1)^2 + (x^2)^2 + (x^3)^2}$, $n^j = x^j/r$ and $dA$ is the surface element.

Here, the spatial component of the metric for large values of $r$ is given by
\begin{equation}
g_{ij} = \left[1 + \frac{2 m}{r}\left(\frac{1 + 2\lambda^2}{1 + \lambda^2}\right) + \mathcal{O}\left(\frac{1}{r^2}\right)\right] \delta_{ij}
\,\,\label{} \, .
\end{equation}
Then 
\begin{equation}
g_{ij,i} - g_{ii,j} = -2 \partial_j \left[\frac{2 m}{r}\left(\frac{1 + 2\lambda^2}{1+\lambda^2}\right) + \mathcal{O}\left(\frac{1}{r^2}\right) \right] 
\,\,\label{} \, .
\end{equation}
Consequently
\begin{equation}
M_{ADM} = \frac{-1 }{8 \pi} \int \partial_r \left[\frac{2 m}{r}\left(\frac{1 + 2\lambda^2}{1+\lambda^2}\right) \right] dA = m\left(\frac{1 + 2\lambda^2}{1+\lambda^2}\right)
\,\,\label{} \, .
\end{equation} 


\section*{Acknowledgements}

We wish to thank the anonymous referee for a constructive criticism a well as for useful comments and suggestions.
A.~R. acknowledges financial support from the Generalitat Valenciana through PROMETEO PROJECT CIPROM/2022/13.
A.~R. is funded by the Mar{\'i}a Zambrano contract ZAMBRANO 21-25 (Spain).
The author L.~B. is supported by DIUFRO through the project: DI22-0026.

\section*{Data Availability Statement}
No Data is associated with the manuscript.






\end{document}